# Network Coding Meets Information-Centric Networking:

## An Architectural Case for Information Dispersion Through Native Network Coding

Marie-José Montpetit◊, Cedric Westphal*, Dirk Trossen▪

| ◊Research Laboratory of Electronics | *Innovation Center | ▪Computer Laboratory |
| Massachusetts Institute of Technology | Huawei | Cambridge University |
| mariejo@mit.edu | cedric.westphal@huawei.com | dirk.trossen@cl.cam.ac.uk |

### ABSTRACT

The focus of user behavior in the Internet has changed over the recent years towards being driven by exchanging and accessing information. Many advances in networking technologies have utilized this change by focusing on the content of an exchange rather than the endpoints exchanging the content. Network coding and information-centric networking are two examples of these technology trends, each being developed largely independent so far. This paper brings these areas together in an evolutionary as well as explorative setting for a new internetworking architecture. We outline opportunities for applying network coding in a novel and performance-enhancing way that could eventually push forward the case for information-centric network itself.

## I. INTRODUCTION

The "Internet of the Future" is already upon us. The models that until a few years ago seemed to describe the behavior of Internet traffic have become obsolete. Indeed, the Internet is less and less about flows moving from a source to a destination but more and more about information disseminated across a large number of nodes. The information super-highway has evolved into a complex organism of information diffusion.

In this context, a re-architecture of the way we consider IP packets becomes necessary. Since the analogy of cars and trucks on the information highway is now inexact in the new very heterogeneous network infrastructure and device ecosystem of today, a new paradigm that follow a more diffusive model needs to be defined. In particular, in information centric networks with multiple rendezvous points, a packet structure that takes advantage of multipath and combination opportunities is needed.

This paper presents such a novel approach. We propose to use network coding (NC), already proven to provide solutions to a variety of networking problems in wireless and wired networks alike, to meet the challenges of information dissemination in content-centric networks (CCNs) as presented in [1]. We want to provide a strategy wherewith the network coding allows for the network traffic to be considered as information not simply bits and that information can be combined, disseminated and extracted within the changing network infrastructure and provide a stateless architecture well suited to the named networks of the near future.

The paper is divided as follows. A short section on network coding basics in Section II is followed by an overview of NC for content dissemination in physical and overlay networks (such as social networks) in Section III. Section IV shows how the approach applies to the so-called CCNs where the IP infrastructure is replaced by a content-centric narrow waist. In Section V, the NC approach is applied to the more revolutionary network paradigm where the current Internet is superseded by a novel non-address based architecture. Section VI provides some views of future work as this our NC-related research is only beginning but opens interesting avenues and challenges for the next years.

## II. A SHORT NETWORK CODING PRIMER

In the Internet today bits are considered to be meaningless. To the contrary, Network Coding (NC) considers data traffic as algebraic information [2]. The output of a network coder is a linear combination of a number of input packets. NC has been shown in an information theoretic manner to reduce the required number of transmissions to complete a file or stream operation over noisy or unreliable networks. NC does add complexity to both source and destination nodes since it involves performing linear operations these are quite simple for the current generation of network elements and end devices.

In order to understand the role that NC can play in disseminating information in a densely connected network, let us consider a simple example (Figure 1). Figure 1 a) shows a traditional way of looking at data dissemination in a multipath network; in order to guarantee the packet delivery they are replicated on two (2) different paths. The goal is to recuperate 3 packets at the receiver. In a) packet 1 is lost on each path due to some path impairment or congestion policy. There is no way to recuperate packet 1 unless there is a feedback loop to the sender and the receiver keeps track of lost packets. In b) linear combinations of the 3 packets are sent on both paths. While the two (2) first packet packets are still lost the 3 packets received contain enough linear combinations, namely three (3) to decode the packets. While this is a simplistic example, it does show that mixing packets increases resiliency and allows implementing a simpler content dissemination approach that does not require to keep track of received packets.

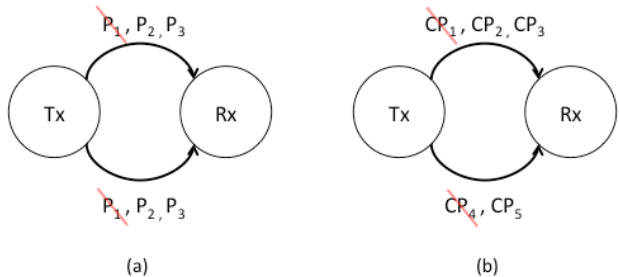

**Figure 1 Multipath Packet Transmission Example a) Without NC and b) with NC.**

In terms of implementation of Network Coding, the use of Random Linear Network Coding or RLNC creates the simplest encoder [3]. In RLNC, the coding coefficients are chosen randomly over the chosen Galois Field, the latter chosen to provide the lowest overhead. However, coding across M packets requires M coded packets to recover any information. In order to make the operation of the decoder simpler, a systematic structure is often used. In this arrangement, the original packets are sent without coding once as if their coding coefficient was 1. All additional packets are sent coded with RLNC [4]. Systematic network coding provides no degradation in performance while ensuring a significant reduction in decoding complexity [4]. It should be noted that NC does not need all packets to be the same size nor does it require to "wait" for a packets of a linear combination to arrive or to be decoded before the next combination is sent.

RLNC has already been used to improve transmission of data over lossy wireless networks by enabling multipath and multi-homing [5][6] in TCP networks. But caching and network coding are getting a lot of attention. As will be discussed in the next sections, it also offers great promise in information centric networks as well as efforts to redefine the Internet hourglass with a content-oriented waist.

III.   CONTENT DISSEMINATION AND NETWORK CODING

The network of today exemplifies the convergence of voice, text, video and gaming. Ancillary content and extra features are combined to the main streams and can be inserted anywhere in the delivery chain. As mentioned in the introduction, the Internet is more and more about content "packages" and collaborative nodes and less and less about data flows. Content-centric network are designed to meet this increase in traffic complexity. When combined with the scarcity of wireless resources and the growing energy use of networks in storing and conveying media one can see that we truly require new architectures. Network coding with its demonstrated resilience to losses and its inherent algebraic structure has already been implement in a content-centric manner[7][8][9].

As the example of the last section demonstrated, in a multi-path network NC naturally lead to better peer-to-peer implementation. In this case information can be shared locally with each node assuming the role of a re-encoder for speed of dissemination without wasting bottleneck resources and without the use of trackers or other means of tracking packets. This approach was used successfully in a number of instances including video distribution in a commercial context[10]and more recently over networks of handheld devices[11]

A natural extension of this approach is to protect the information to provide both privacy and protection against rogue transmission[12]. The combination of peer-to-peer and privacy naturally leads to a very secure dissemination network that reduces the reliance on bottleneck resources for non-revenue generating traffic while allowing peer to peer transmission of content at the edge [13].

Finally, if we consider the new network as a mesh of local caches to serve both mobile and fixed users then the use of NC segments in the caches allows better management of the file access with reduced blocking. This can work with current caching policies as well as allow to define new ones such as mechanisms to treat popular offerings differently from long tail content. While this is work in progress, applications are envisaged including the distribution of video content in mobile and vehicular networks.

While the benefits of the use of network coding in content dissemination has been established above in some p2p or content overlays, we contend that it can –and indeed should- be included in the networking layer architecture for future information-centric networks. The next two sections consider how to include network coding into the networking layer of ICNs, by considering two potential such architectures: first, we consider a CCN-based approach [14] in the next Section, and a pub/sub mechanism in Section V.

IV.   NETWORK CODING IN THE CCN ARCHITECTURE

CCN as defined in [14] proposes to replace the IP narrow waist of the Internet architecture by a content layer. We now briefly describe how CCN works.

When a node wants to access a piece of content, it sends an *Interest* packet to the network. The network then, using the name of the content for routing, forwards the Interest to one or more copies of the content object.

Once the Interest reaches a cache holding a copy of the content object, a Data packet is sent back. The Data packet retraces the path followed by the Interest in the reverse direction to the node which requested the content. Each Data packet is a chunk of a larger content object. The size of each chunk is not fully specified in this architecture, but should be chosen to optimize the trade-off between various parameters: maximum transfer units on the intermediate links, volume of the Interest traffic and latency at the receiver.

The Interest can request a specific chunk, say "www.foo.com/Dir/File/C1" or just initiate the transfer of file by requesting "www.foo.com/Dir/File/" which is implicitly understood to send the first chunk in return.

This architecture naturally allows routers along the path to store the content in its local storage, denoted as the Content Store. This allows the router to serve the cached content in response to an Interest rather than going to the origin server of the content.

One key aspect of CCN is to fully disconnect the delivery of the content from any network location: rather than establishing an end-to-end connection, the content is received chunk by chunk from wherever those chunks were stored in the network. This is also embodied within the security architecture, which is built independently of the connection endpoints.

We now describe why network coding would bring significant improvement to this specific architecture

### A. Inherent Multipath Support

CCN does not tie up the exchange of data to a single interface. Indeed, there is no need to establish a connection at the network layer to initiate a request for content. This in turn implies that a node can send Interest to several interfaces (say, 3G and WiFi), receive the data from these multiple interfaces, and recompose the content object for application use.

This falls squarely in the illustration of Figure 2. The node could forward Interest packets for an object composed of two chunks C1 and C2 over two different interfaces, say 3G and WiFi. Without network coding, the Interests would get to copies of the object, and start the transmission of the first chunk C1. The node would thus receive two copies of C1 from each interface (provided that the paths are fully independent).

On the other hand, with network coding, rather than sending C1 as an implicit response to "www.foo.com/Dir/File", both cache servers would send a linear combination of the two chunks, say C1+2C2 for the first one, and 2C1+C2 for the second one. Alternatively, the name could explicitly make a request for network coded chunks, say using a specific syntax such as "www.foo.com/Dir/File/NCChunk".

Upon reception of these two encoded chunks, the node can reconstruct the full data objects C1 and C2. Note that the transmission bandwidth used in the case of network coding is exactly the same as in the case without, but in the former case, the whole data object was retrieved, while in the second one, half of the capacity was wasted.

Note also that this was achieved in an asynchronous manner. The only requirement is that the encoding of the packets received from the two caches be linearly independent. Since these are generated independently, it might not always be the case, but for randomly generated codes, the probability of such an occurrence can be made arbitrarily low.

Finally, observe that the rate of each transmission over the two different interfaces does not have to be known a priori. If one interface was three times faster than the other, three times as many linear combinations would arrive there, but both interfaces would be bringing fresh information at their respective full throughput. The total transmission rate would be the additive capacity of both interfaces, and this capacity would be achieved in a fully distributed manner.

### B. Caching in CCN

Caching is inherently supported in the CCN architecture. Any router can, if it supports this, cache the content for further use. In the CCN architecture, the router also aggregates the Interests and consolidates the transmission of Data packets. A router, if it does not possess a route to a specific object, can also broadcast an Interest over multiple interfaces in order to locate the object.

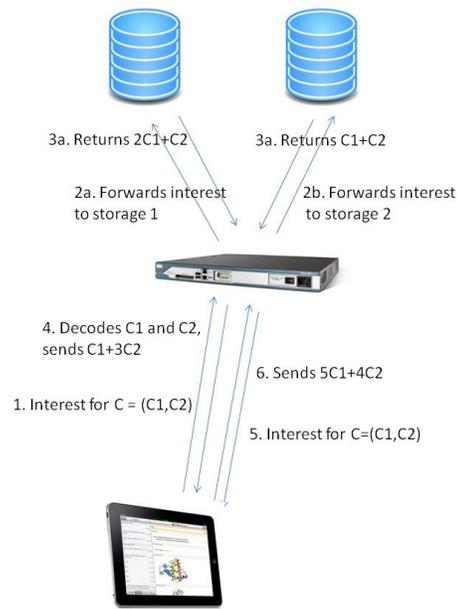

**Figure 2 Network-Coded Forwarding**

As an example, consider the same object composed of chunks C1 and C2 and consider a router R which attempts to locate this object upon receiving an Interest from node N. The router may not have an entry for this name in its Forwarding Information Base and might attempt to find it from two neighbours. If both these neighbours do have the content, they both will respond to the Interest with C1.

The router will discard one of the copies, and respond to the Interest with C1. However, N will then issue an interest for C2. At this point, R will fetch C2 from one of the now known repositories[1].

In the case with network coding, R would have received two linear combinations after forwarding the first Interest from N to its neighbours. From these two linear combinations, it would have forwarded one of them to N and upon receiving the second interest, the other[2].

In this case, the total bandwidth amount consumed is reduced as in the previous example, but the delay is also reduced by taking advantage of the caching opportunity within the network. The second roundtrip to fetch C2 is only from N to R, and not all the way to one of the repositories.

In CCN without network coding, caching would benefit another user N' when she requests C1 and C2 from R later on. With network coding, it also benefits N during the first transmission by populating the cache at R faster.

The examples we describe use network coding to encode within the same object transmission as in [6]; of course, if a router receives different linear combinations intended for different receivers, it can combine these as well in the same manner.

*C. NC3N*

We now describe in a more generic fashion our Network Coding for CCN (NC3N) architecture proposal.

One key observation is that each chunk in CCN carries already some metadata; in particular, it carries some security information regarding the content of the Data packet.

Our proposal is to simply add one field each into the Interest packet and the Data packet semantics. The header of the Interest packet contains three fields: Content Name, Selector and Nonce. We suggest inserting in the Selector a flag allowing the transmission of network coded chunks in response to the Interest carrying this flag.

This flag would be turned on for instance in any case where multiple packets could be received in response to the interest (say, because it is broadcasted to several neighbours, or sent over multiple interfaces). The flag also could be leveraged in the caching policy, to decide whether or not to cache encoded chunks.

The flag is required as some data exchanges require the packet to be received in a specific order, for instance to start streaming a video before all chunks are received[3].

A Data packet issued in response of such an Interest would carry a modified field. Currently, the structure of a Data packet contains the Content Name, a Signature, some Signed Information and the Data. The Signed Info would need to carry the coefficient of the linear combinations, and the Data would carry the encoded object thus described in the Signed Info.

These integers should be chosen randomly from a set in order to avoid linearly dependent combinations to be generated at different nodes.

The operation of NC3N works simply as follows: the sender of the Interest, if it supports network coding, sets the flag up in the Interest. Upon forwarding a response to the interest, each node would look up the number of coefficients and generate an encoded version. Each response to an Interest generates a new encoded version, so that the receiver of the content gets new degrees of freedom with each chunk it gets. Intermediate nodes can cache the encoded chunk. If it holds several such chunks, it can also generate new combinations. If it holds enough chunks to decode, then it does so. An intermediate node should respond to an Interest only if it can provide new degrees of freedom from those mentioned in the Interest Selector.

NC3N is a relatively straightforward evolution of CCN, but would bring significant benefits in bandwidth reduction and delay. The overhead is relatively minimal, as the cost is only one bit in the Selector if it is used to turn off network coding, and a few bits in Interest and Data packets when it is turned on (presumably because of an anticipated benefit).

## V. NETWORK CODING FOR A PUB/SUB ICN ARCHITECTURE

Let us now present the potential for network coding within an approach for changing the internetworking architecture towards a publish/subscribe information-centric layer. We base our discussion here on the architecture presented in [1]. In this work, complemented by project efforts such as in [15], the authors envision a rendezvous sub-system to provide a late location binding functionality, matching information availability to interest at runtime. A separate topology management function determines a suitable communication relation between the providers and consumers of information, with a forwarding function eventually delivering the information throughout the network. Key here is the uniform positioning of *information* as the main principle of the architecture; a principle that aligns well with the ideas of network coding. In the following, we elaborate on three potential use cases for native network coding in such more revolutionary setting.

*A. Rendezvous*

The process of rendezvous requires *matching* providers and consumers for information among a possible large set of candidates. For this, the rendezvous function provides a registration process for availability of as well as interest in information, such registration realized via, e.g., publish/subscribe interface semantics. The realization of the rendezvous function ranges from centralized, e.g., domain-local, servers over hierarchical DHT approaches [18] to fully distributed solutions. Let us focus on the latter as an example for utilizing network coding.

---

[1] This assumes the cache hold the whole objects, not individual chunks. This depends on the caching policy which is not specified in CCN. Note that if caches hold only single chunks and not whole content objects, then having network coding increases the likelihood of finding the missing bits of content.

[2] Or it could have decoded C1 and C2 and sent C1 in response of the first interest, and C2 for the second.

[3] In this case, the network coding can also be applied to small groups of chunks, say the first $k$ chunks, then the next $k$, etc. The stream only needs to receive $k$ chunks before starting, which is similar to the buffering mechanisms in current video players.

The first example is that of dispersing the matching requests themselves. Within a fully distributed rendezvous function, the matching requests are distributed to the 'best' candidate that could potentially match the request. For such distribution of requests, a solution could utilize network coding solutions similar to [10] to disperse rendezvous requests among a set of rendezvous points that serve a particular domain (or part of the information space the rendezvous points as a whole serve). For this, the requests are treated similar to encoded video content in [10]. Crucial in such solution, however, are the timing constraints for the rendezvous requests. With rendezvous being part of the overall control path of the final data delivery, a dispersion-based mechanism might not yield the necessary performance in terms of delay. On the other hand, in scenarios where the rendezvous function is only initially invoked with many information transfers following the original match[4], the possible advantages of the network coded dispersion, such as resilience, might outperform any possible delay penalty during the initial setup.

Our second example relates to the exchange of rendezvous state rather than the matching requests themselves. Here, we address the problem of synchronizing the information required within each rendezvous point to perform the desired matching. Such information includes the structure of the information space, policies attached to (parts of) the information space as well as information about providers and consumers. While we can assume the latter to be possibly partitioned across the distributed rendezvous points, some form of common state on the former, i.e., the information structure in which the matching is performed, is required. Dispersing this information between the individual rendezvous points is similar to a secure file exchange in a traditional overlay network coding example of today's Internet and solutions such as proposed in [12] can be applied in this context. Hence, it almost naturally lends itself to using network coding between the rendezvous points.

### B. Cache Replication and Management

Our second use case for applying network coding as part of the network infrastructure is that of cache replication and management. While approaches like CCN[14] foresee caching to occur on a per-router basis, we see this form of (transient) caching being supplemented by *managed caches* that hold replicated content according to some policy, such as business contracts, popularity or local relevance. Within the architecture, these caches serve as additional publishers for the requested (managed) content. We expect the content of these caches to be replicated on a mid-term time scale, typically within days (see [16] for a replicated cache solution within an information-centric network setting).

It is this replication where network coding can be naturally applied as the basic information forwarding policy. For this, the sub-space of the information structure that is planned to be cached would be network encoded and forwarded using the basic forwarding policy of the domain in which the managed caches reside[5].

### C. Forwarding

Our third example is that of extending the fast-path forwarding policies for information-centric solutions with network coding.

We base our discussion here on the forwarding solution outlined in [17]. This mechanism provides native multicast forwarding of packets by virtue of a constant length identifier that encodes the overall multicast tree using a *Bloom* filter. Although the usage of a constant-length Bloom filter counters the problem of growing path lists in usual source-based routing approaches, it comes with the drawback that the likelihood of false positive forwarding decisions increases with the length of the encoded multicast tree. The original solution presents workarounds for this issue[17]by essentially 'thinning out' the increasingly crowded link identifier space that needs to be folded into the Bloom filter. It is here that Network coding can further alleviate the problems of false positives by coding across separate sessions of otherwise disparate information flows.

Let us illustrate this with a simple example of two pub/sub relations for information A and B, as shown in Figure 3. Assume that either one or both information flows suffer from an inappropriate rate of false positives in the respective Bloom filters. Let us further assume that the delivery graphs partially overlap. In this case, one can construct a third delivery graph for the network coded information flow A+B. Each of the forwarding nodes at the border of the common partition needs to be NC-enabled. An *incoming* NC-enabled forwarding node (a) network-encodes the information flows A and B (b) creates a new information flow A+B with an *algorithmically derived*[6] identifier and (c) forwards the information according to the forwarding identifier for A+B (which it adds to the encoded packet for further forwarding). An outgoing NC-enabled forwarding node reverses the operations, i.e., decapsulates the information flow A+B (by removing the forwarding identifiers A+B), decodes the individual flows and restores the individual forwarding identifiers for A and B (obtained through the decapsulation).

---

[4] Examples for such scenarios are video streaming or sensor push scenarios. Here, the initial location binding, e.g., between a video provider and its consumers, is followed by the transfer of many information items; the information semantics here is that similar to a *channel* of information rather than that of a *versioned document*.

[5] The range of these forwarding policies could span from local broadcast to Bloom filter based forwarding as outlined in the following Section XX.C.

[6]*Algorithmically identified* refers to a technique where the identifier of a new flow is algorithmically related to the ones that are used for its construction. Such algorithmic relation can be a hash or tree algorithm.

Such integration of network coding allows for various decision criteria as to how to construct appropriate subgraphs. Apart from the false positive rate, other criteria could include congestion on a given link (such as the first A+B link in Figure 3), triggering to partially use network coding on this segment in order to increase the overall throughput. Another criteria could be resilience for a given stream by coding stream A into another stream's subgraph, utilizing the additional route taken for resilience purposes.

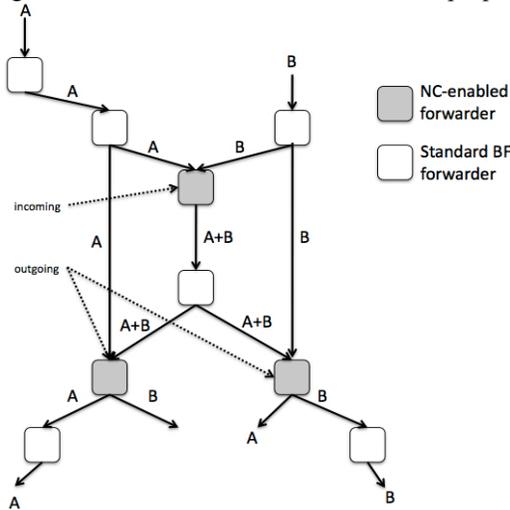

**Figure 3: Network-Coded Forwarding**

The support for network coding comes, however, with a price. Firstly, there is the required support for encapsulation and network coding, as outlined above. In addition, the information flow would need support for identifying individual fragments that are appropriately re-assembled in the NC-enabled nodes. Only a thorough evaluation of this approach will shed light on the appropriateness of such extension. However, we foresee that even partial support in a few nodes could lead to significant improvements for reducing false positives and easing congestion.

## VI. CONCLUSIONS

Network coding has been identified as an area of tremendous potential in wireless and wired networks alike for quite some time now. In this paper, we attempted to push the vision of network coding beyond its current usages within overlays for content dissemination.

For that, we outlined potential application areas for network coding within content-centric approaches to networking that focus on information as the main principle of interaction. This has led us to identify concrete examples where network coding could greatly enhance the overall performance in a CCN setting as well as in pub/sub approaches.

We clearly recognize that providing comprehensive evidence for the claimed benefits is the next big step in our future work. But it is the identification of the opportunities itself that we see as potentially stimulating an increased interest in pushing forward the vision of network coding, specifically within the area of information-centric networking.

## VII. ACKNOWLEDGEMENTS

We would like to acknowledge Muriel Medard for her contribution to the overall framing of the problem space and Gergely Biczok for his contribution to utilizing networking coding for information-centric forwarding.

Marie-Jose's work was partially supported by FT/Orange R&D. Dirk's work was funded by the EU FP7 project PURSUIT under grant no. FP7-INFSO-ICT-257217.

REFERENCES

[1] D. Trossen, M. Sarela, K. Sollins, "Arguments for an Information-Centric Internetworking Architecture", ACM Computer Communication Review 40(2), pp. 27-33, 2010.

[2] R. Koetter, M. Médard, "An Algebraic Approach to Network Coding." IEEE/ACM Trans. on Networking, Vol. 11, No 5, 2003.

[3] T. Ho, R. Koetter, M. Médard, D. R. Karger and M. Effros, "The Benefits of Coding over Routing in a Randomized Setting". IEEE International Symposium on Information Theory, 2003.

[4] D. E. Lucani, M. Médard, M. Stojanovic, "Systematic network coding for time-division duplexing." Proc. IEEE International Symposium on Information Theory. pp.2403-07, Austin, TX, 2010.

[5] A. Kulkarni, M. Heindlmaier, D. Traskov, M. Medard, M.J. Montpetit, "Network Coding with Association Policies in Heterogeneous Networks", NC-Pro 2001, May 2011.

[6] J. K. Sundararajan, D. Shah, M. Médard, S. Jakubczak, M. Mitzenmacher, J. Barros "Network Coding Meets TCP: Theory and Implementation", Proceedings of the IEEE, March 2011.

[7] A. Dimakis, P. Godfrey, Y. Wu, M. Wainwright, K. Ramchandran, "Network Coding for Distributed Storage Systems," IEEE Trans. Information Theory, vol. 56, No. 9, September 2010.

[8] H. Seferoglu, A. Markopoulou, "Opportunistic Network Coding for Video Streaming over Wireless", in Proc. Packet Video'07, Lausanne, CH, November 2007.

[9] C. Gkantsidis, P. Rodriguez, "Network Coding for Large Scale Content Distribution," in Proc. of IEEE Infocom'05, Miami, FL, March 2005.

[10] Z. Liu, C. Wu, B. Li, S. Zhao, "UUSee: Large-Scale Operational On-Demand. Streaming with Random Network Coding." Infocom 2010.

[11] P. Vingelmann, F.H.P. Fitzek, M.V. Pedersen, J. Heide, and H. Charaf, "Synchronized multimedia streaming on the iphone platformwith network coding," in IEEE Consumer Communications and Networking (CCNC), Las Vegas, NV, USA, Jan. 2011.

[12] L. Lima, S. Gheorghiu, J. Barros, M. Médard, A. T. Toledo, "Secure Network Coding for Multi-Resolution Wireless Video Streaming," IEEE JSAC, Vol. 28 No. 3, April 2010.

[13] M.J. Montpetit and M. Médard, "Community Viewing meets Network Coding: New Strategies for Distribution, Consumption and Protection of TV Content", Presented to the 2nd W3C Web and TV Workshop, February 2011.

[14] V. Jacobson, D. Smetters, J. Thornton, M. Plass, N. Briggs, R. Braynard, "Networking Named Context", In: Proc. of ACM CoNext'09, December 2009.

[15] PURSUIT project, available at http://www.fp7-pursuit.eu, 2011

[16] P. Flegkas, V. Sourlas, G. Parisis, D. Trossen, "Storage Replication in Information-Centric Networking", Proc. of IEEE ICON, 2011

[17] P. Jokela, A. Zahemszky, S. Arianfar, P. Nikander, C. Esteve, "LIPSIN: Line speed Publish/Subscribe Inter- Networking", ACM SIGCOMM, August 2009.

[18] J. Rajahalme, M. Särelä, K. Visala, J. Riihijarvi, "On name-based inter-domain routing", Computer Networks 55:975–986, 2011.